\documentclass[article]{aa}
\usepackage{psfig,latexsym,longtable,lscape,supertabular}

\def\e20{$\times 10^{20}$}
\def\ergsec{erg s$^{-1}$}

\def\chandra{{\it Chandra}}
\def\lx{L$_x$}

\begin{document}

\pagenumbering{arabic}
\title{The peculiar small-scale X-ray morphology of NGC 5846 observed
with {\it Chandra}}

\author{Ginevra Trinchieri\inst{1} \and Paul Goudfrooij\inst{2}}
\institute{
Osservatorio Astronomico di Brera, via Brera 28, 20121
 Milano, Italy
\and 
Space Telescope Science Institute, 3700 San Martin Drive, Baltimore,
 MD 21218, U.S.A.
}
   \offprints{G.~Trinchieri}
   \mail{ginevra@brera.mi.astro.it}

   \date{Received date; accepted date}
\titlerunning{
Small-scale X-ray morphology of NGC 5846}
 
\abstract{
The excellent quality of the \chandra\ observation of NGC 5846
reveals a 
complex X-ray morphology of the central regions of this 
galaxy. An intriguing morphological similarity between the X-ray and the
optical line emission, discovered before using ROSAT HRI images (Trinchieri,
Noris \& Di Serego Alighieri 1997), is confirmed here in unprecedented detail. 
Complex spectral characteristics are associated with the morphological
peculiarities, indicating a possibly turbulent gas in this object. 
A population of $\sim 40$ individual sources is also observed, 
with \lx\ in the range $\sim 3 \times
10^{38}- 2 \times 10^{39}$ \ergsec, with an X-ray luminosity function
that is steeper in the high-luminosity end than in other early-type
galaxies.
\keywords{Galaxies: individual: NGC~5846 --
		X-rays: galaxies}
}

\maketitle

\section{Introduction}
The hot interstellar medium (ISM) in bright early-type galaxies
was first imaged with data from the {\it Einstein\/} Observatory, and a
few of its broad characteristics have been established in the more
recent past mainly with observations from ROSAT and ASCA.  Thermal
emission from early type galaxies with a large $\rm L_x/L_B$ ratio is
characterized by an optically thin plasma spectrum,  with typical
average temperatures of $\sim 0.5-1$ keV (Forman, Jones \& Tucker 1985;
Canizares, Fabbiano \& Trinchieri 1987) and extent that could vary from
galactic scale out to several hundred kpc.  The hot ISM is not
isothermal over the entire source, but can show a slow temperature
increase over a large region (eg, $\sim$ 40 kpc in NGC~4636, Trinchieri
et al 1994), and its large scale morphology can either be relatively
smooth and regular, or  show asymmetries and
inhomogeneities (eg., the characteristic``tail" in M86, Forman et al. 1979;
also less dramatic distortions are found in other galaxies, see
Trinchieri et al. 1994; 1997b; Kim \& Fabbiano 1995; and references
therein).   Several studies have focused on better understanding
the hot gas properties in more details, from the relation with the
optical properties, galaxy density and grouping, to the chemical gas
composition, all contributing to a vivacious and lively, though often
controvertial,  debate.  On the
other hand, the small-scale morphology and detailed characteristics of
the ISM  have so far been very poorly studied due to the limited
spatial resolution of the data available prior to the launch of the
\chandra\ satellite and its instruments. 

Coexisting with the hot ISM, a cooler (``warm'') component has been
mapped in H$\alpha$ in several X-ray bright early-type galaxies (Kim 1989;
Trinchieri \& Di Serego 1991; Shields 1991; Buson
et al.\ 1993; Goudfrooij et al.\ 1994; Singh et al.\ 1995; Macchetto et
al.\ 1996).  The morphology of 
this emission can be either relatively smooth and azimuthally symmetric
or exhibit filamentary structures, arcs and/or rings.   Although clearly
extended, it is generally confined to the innermost regions of the
galaxies, often within $\sim$\,1 kpc, and it is relatively faint and hard
to detect properly.

To date, the study of the relation between hotter and cooler gas has
been restricted to a handful of galaxies. In particular, a very
intriguing morphological similarity between X-ray and H$\alpha$
emission has only been established in two normal galaxies not at the
center of clusters, using ROSAT HRI observations (Trinchieri \& Noris
1995; Trinchieri et al.\ 1997a). These studies have shown that the
highly asymmetric, clumpy and filamentary H$\alpha$ emission in NGC
1553 and NGC 5846 correspond to a complex, structured, clumpy emission
in the X-ray images, with gross similarities and some remarkably
similar details as well.  In NGC 4649 however, where the H$\alpha$
emission is smooth and azimuthally symmetric, no structure or
asymmetries are found in the X-ray emission.

Another intriguing suggestion from the X-ray data of early-type
galaxies is the fact that these X-ray features may have different
spectral characteristics. Global spectral variations with radius have
been measured with the ROSAT PSPC, and indicate a variation of
temperature, increasing towards larger radii. However, these variations
are generally azimuthally averaged quantities derived on scales of
arcminutes, and reliable information on small scales could typically
not be derived from the X-ray instruments available up to now.  An
attempt at measuring a spectral variation in a small region was done
using the PSPC observation of NGC 5846 (Trinchieri et al.\ 1997a):  the
gas in the region corresponding to a feature to the NE seemed to have a
cooler temperature than the surrounding gas at the same galactocentric
distance, indicating a cooler plasma in correspondence to enhanced
emission.

\begin{figure}
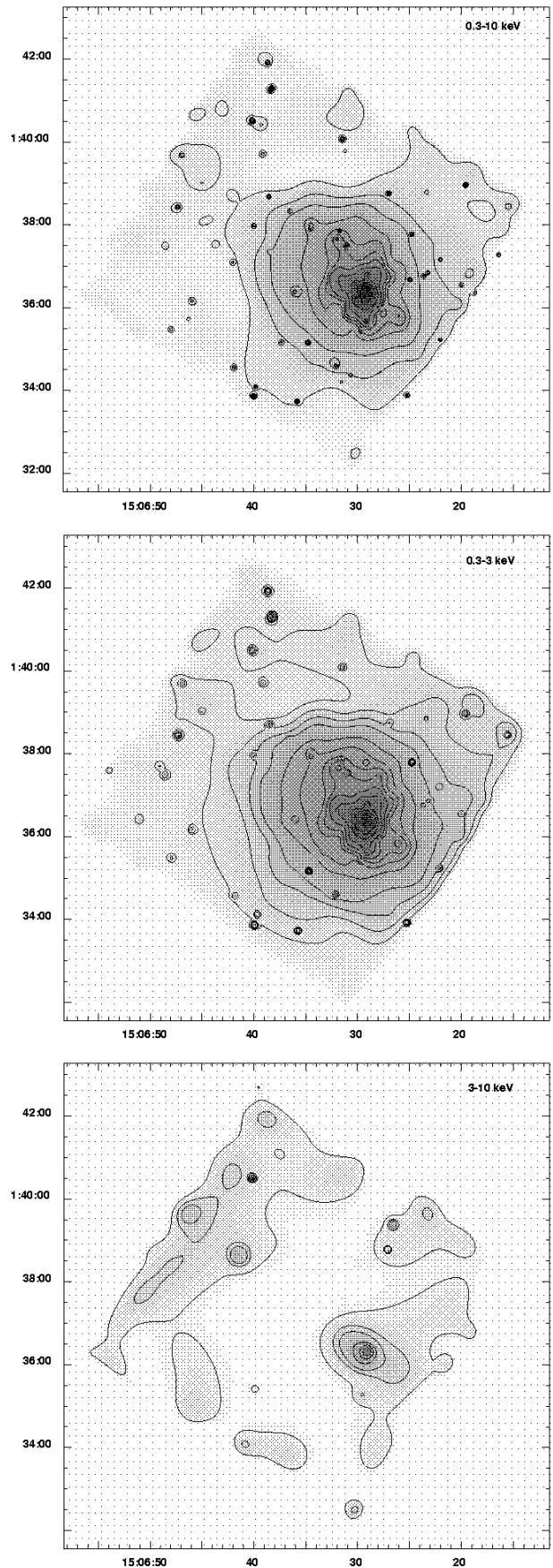

\psfig{figure=2278.f1a,width=8.5cm,clip=}
\psfig{figure=2278.f1b,width=8.5cm,clip=}
\psfig{figure=2278.f1c,width=8.5cm,clip=}
\caption{Isointensity contour plot of the X--ray emission in NGC 5846
in the full energy band (0.3-10 keV), soft (0.3-3 keV) and hard (3-10 keV).  
The data have been smoothed with an adaptive technique (fft method, with
sigmin=2.5 and sigmax=5).
}
\label{large}
\end{figure}

The close similarity of the filamentary structures in X-ray-- and
optical emission in some early-type galaxies, together with the {\it
lack\/} of them at both wavelengths in others, re-enforced the case for
a physical connection between the different phases of the interstellar
medium.  Such a connection had been dismissed within the ``cooling
flow'' model, since predicted H$\alpha$ luminosities obtained from the
cooling gas were significantly smaller than those actually observed
(e.g., Baum 1992).  Given the short cooling times of the X-ray-emitting
gas, a plausible suggestion that has been made is that the hot gas is
directly responsible for line emission through heat transfer from the
hot gas to cooler gas through electron conduction (Sparks et al.\ 1989;
de Jong et al.\ 1990). This scenario indeed explains in detail the
observed enhancement of X-ray emission at the location of dusty optical
filaments. In this respect, it is interesting to note that a
filamentary dust lane in the central few kpc of NGC 5846 has been
detected, with a morphology once again strikingly similar to that
observed for the optical nebulosity {\it and\/} the X-ray emission
(Goudfrooij \& Trinchieri 1998).

The high spatial and spectral resolution, sensitive data set that we
have obtained using \chandra\ will be used in what follows to confirm the
tentative results obtained with the ROSAT HRI and to better investigate
the physical connection between different phases of the ISM in NGC 5846.
This is a giant elliptical galaxy at the center of a small group of
galaxies.  Extended X-ray emission has been detected from the galaxy and
the group already with the $Einstein$ satellite (Biermann, Kronberg
\& Schmutzler 1989) and subsequently studied in detail with the ROSAT 
PSPC and ASCA (Finoguenov et al.\ 1999). 
We assume a distance of 31.6 Mpc, which gives a total L$_x
(0.2-2\,\mbox{keV}) = 6 \times 10^{41}$ \ergsec\ 
and a total gas mass M$\rm _{gas}$ $\sim 10
\times ^{11}$ M\sun\  within 10$'$ radius (from Finoguenov et al.\ 1999).

\section{Results of the \chandra\ data analysis}

\begin{figure*}
\resizebox{18cm}{!}
{
\psfig{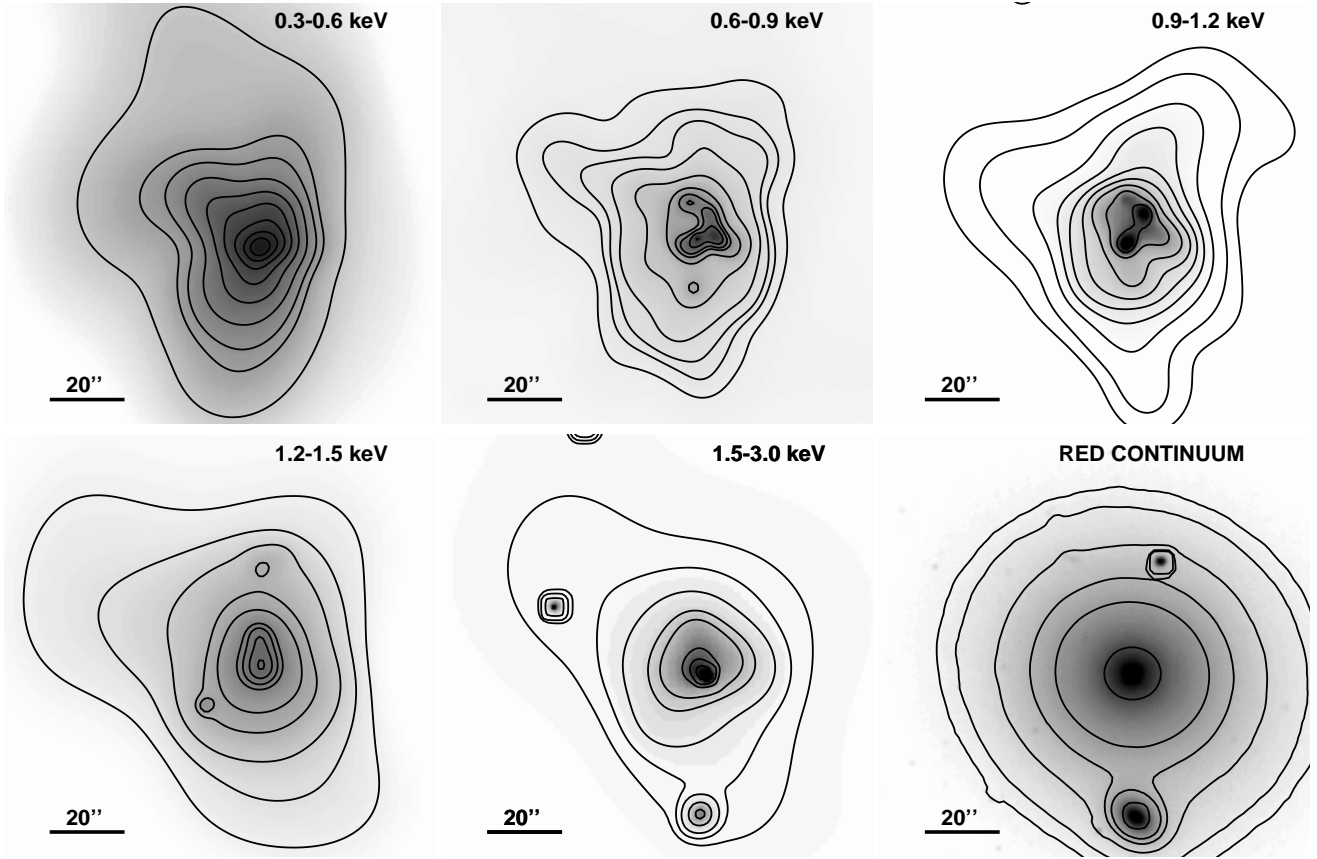}
}
\caption{
Isointensity contour plots of the X-ray emission in the central region
of NGC~5846.  
The different panels represent the data in 
energy bands within the $0.3-3$ keV range, binned to $1'' \times
1''$ pixels and adaptively smoothed (fft kernel).  
The red optical continuum is shown in the bottom right panel on the same
scale. Red continuum data were obtained with EFOSC1 at the ESO 3.6-m telescope
(see Trinchieri \& di Serego Alighieri 1991).  
}
\label{manye}
\end{figure*}

We obtained an observation of NGC~5846 with the back-illuminated S3
CCD (ACIS-S in imaging configuration) on board \chandra\ on May 24,
2000 for $\sim$\,30 ks.  The data have been reprocessed using the new
calibration files as explained in the CIAO documentation, and have been
analyzed with both CIAO\footnote{{\tt http://asc.harvard.edu/ciao/}}
and the {\tt funtool/ds9} software\footnote{{\tt
http://hea-www.harvard.edu/RD/index.html}}.

We have followed the ``CIAO Science Threads" given on the ASC home page
to check and to improve the quality of the data obtained with the
standard processing.  In particular we have eliminated $\sim 5$ ks of
data affected by a high background level, obtaining a net exposure time
of 24264 sec.

Figure~\ref{large} presents the isocontours of the smoothed X--ray image
of the whole back-illuminated CCD in the full energy
range 0.3-10 keV.  The raw data have been smoothed with an adaptive
filtering technique (``csmooth").   
The X-ray emission from this object fills the entire CCD chip, and we
expect it to extend outside, from previous X-ray images of the object in
the soft energy band (cf. PSPC
image, Trinchieri et al.\  1997a; Finoguenov et al.\ 1999).   
However, since we are interested
mostly on the detailed structure present in the X-ray morphology, and
on a detailed study of the characteristics of the emission at small and
medium scales, we will not extend the data analysis outside of the
single back-illuminated S3 CCD.

\begin{figure}
\psfig{file=2278.f3a,width=9cm,clip=}
\psfig{file=2278.f3b,width=9cm,clip=}
\caption{Radial profile of the total (top) and net(bottom) 
emission in the energy bands indicated.  The dashed line represents the
profile in the full 0.3-10 keV energy band.
The profiles have been arbitrarily rescaled : to the background level in
the top panel; to the peak value in the bottom one.  
Harder profiles are obtained in larger annuli than softer profiles, to
retain a better statistics. 
The net surface brightness in the hardest band ($>$ 3 keV) is confined in
the inner $1'$ region and is not shown. 
Errors are indicated for the one band only (1.5-3 keV) only for clarity.
}
\label{rawprof}
\end{figure}

\subsection{Maps}
\label{maps}
Two broad energy band images 
($0.3-3$ keV and $3-10$ keV) 
are also shown in Fig.~\ref{large}.  It is
quite clear from the comparison of the two maps that the 
emission is mostly confined below $2-3$ keV, and that it has a 
general elongation to the NE and a rather complex morphology in
the inner regions. 
Several small blobs can also be 
seen. These 
are most likely individual
point sources, as will be discussed later (see Sect.~\ref{individual}).
Several of these are also seen at 
higher energies, while the more diffuse component is almost entirely
gone above $\sim$\,3 keV.  

Close-ups of the inner region in different energy bands 
show more clearly the complexity of the morphology
at small scales (Fig.~\ref{manye}). 
Strong deviations from azimuthal symmetry are evident 
in all energy bands, 
but the details of the pattern are not the same at all energies. 
The comparison with the central region of the galaxy in the optical is
also quite striking, since none of the perturbations evident in the
X-ray morphology is visible in the stellar light distribution.

The main feature of the X-ray emission is an arc-like or ``hook" structure
at the very center 
which is 
composed of ``blobs'' of emission at slightly different temperatures 
(or, more precise, 
with a different peak and shape in the photon distribution).  A more
coherent structure is visible at $> 1'$ distance from the central X-ray
peak, with an excess  mostly to the NE, that again has no 
counterpart 
in the stellar light distribution.  At hard X rays (kT $\geq$ 3 keV), only
the individual sources remain, and the coherent though distorted 
structure is no longer visible.

\subsection{Radial profiles}
\label{profiles}
\begin{figure*}
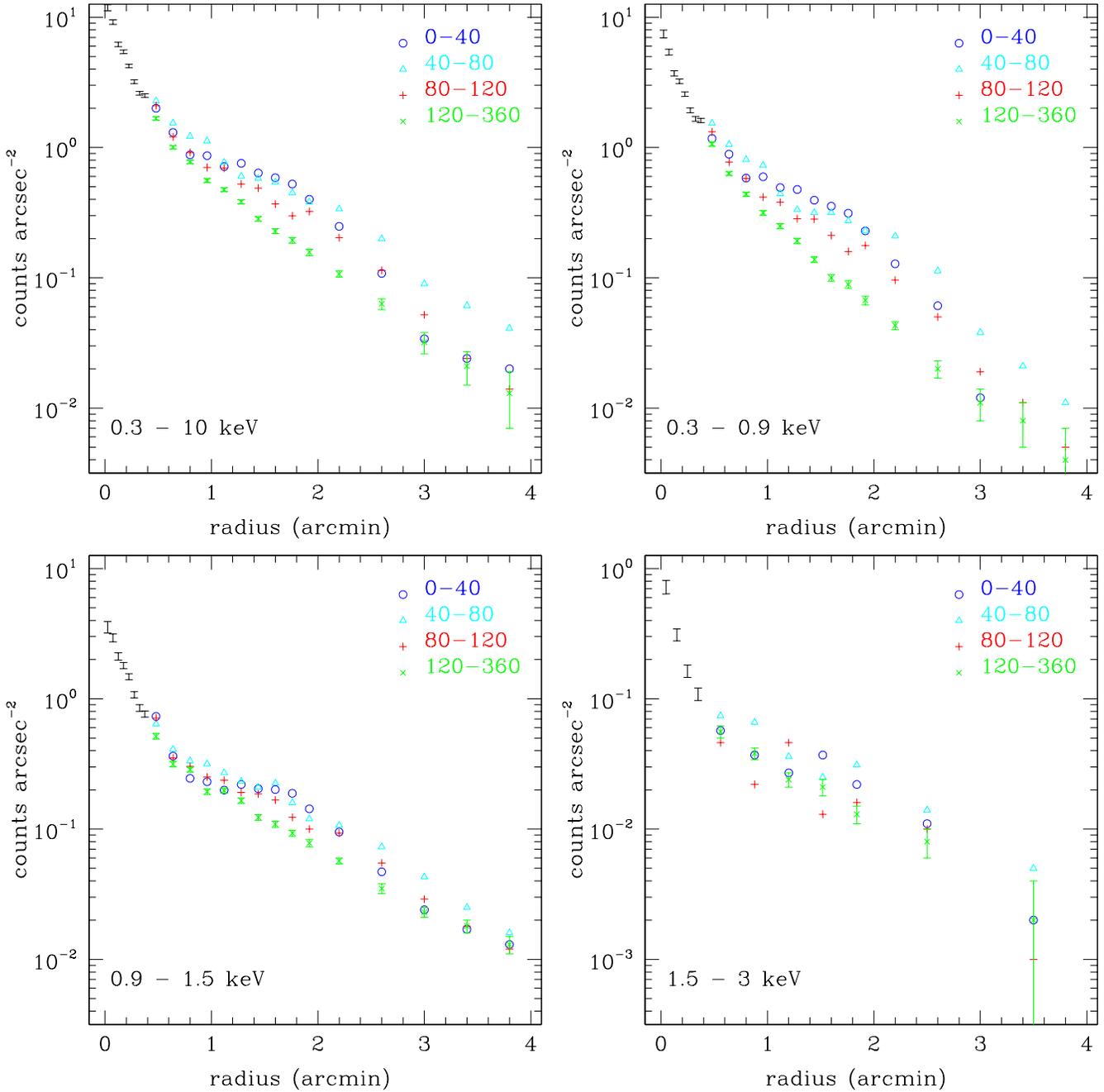

\resizebox{18cm}{!}{
\psfig{file=2278.f4a,width=9cm,clip=}
\psfig{file=2278.f4b,width=9cm,clip=}
}
\resizebox{18cm}{!}{
\psfig{file=2278.f4c,width=9cm,clip=}
\psfig{file=2278.f4d,width=9cm,clip=}
}
\caption{Radial profile of the net emission in different azimuthal
quadrants, as indicated.  
Top left: $0.3-10$ keV band; Top right: $0.3-0.9$ keV; Bottom Left: $0.9-1.5$ 
keV; Bottom right: $1.5-3.0$ keV.
The profiles are derived out to r$\sim 4'$, but in the 120\degr--360\degr\
region the area covered by the CCD field of view is much smaller than in
the Eastern profiles, due to the off-center position of the source. 
}
\label{angprof}
\end{figure*}

We have derived azimuthally averaged radial profiles in several energy
bands and in several azimuthal sectors, as shown in Fig.~\ref{rawprof}
and Fig.~\ref{angprof}.
We centered the profiles at 
(RA, DEC) = (15$^h$06$^m$28\fs911, +01$^d36'$22\farcs35).  
The choice of the center is not 
straightforward, since there is no clearly defined peak common to
all energies (see Fig.~\ref{manye}).  However, since the
central region is highly complex, we
will disregard the innermost points for now.  All 
point 
sources detected (see
Sect.~\ref{individual}) have been excluded from the profile. 
Since the  target is  not located at the center  of the CCD, 
we have excluded regions not covered by the CCD
by selecting photons within the largest rotated box consistent with the
size and orientation of the chip.  
Therefore, the radial profile for radii larger than $\sim 2'$ 
cover progressively smaller regions of the CCD.  

In Fig.~\ref{rawprof} we show the radial profiles of the raw counts,
azimuthally averaged over 360\degr\ and in 4 broad energy bands.  Only
at radii $r >4'-5'$ are the profiles constant with radius in all bands,
indicating that emission is present over the almost entire field of
view.   At hard energies ($> 3$ keV), the profile appears to flatten at
considerably smaller radii.  To estimate the ``field" background for
this observation, 
consistently 
for all energies, we have chosen a
small region in the E corner, at a radius of $\ga 5'$ from the peak,
where the surface brightness appears constant at all energy bands.   In
the total band, the value of the assumed background is $\sim 3.75
\times 10^{-6 }$ cts s$^{-1}$ arcsec$^{-2}$, higher than the expected
ACIS-S background of $\sim 3 \times 10^{-6}$cts s$^{-1}$ arcsec$^{-2}$
(from 
blank sky fields).   This could be due to background
fluctuation in different sky regions, most likely with a contribution
from the galaxy itself, that is known to extend well beyond the area
covered by the S3 chip.

The radial distribution of the emission is rather complex, and
appears to have different
radial gradients, distribution and extent both as a function of energy
(see Fig.~\ref{rawprof}) and as a function of angular sector
(Fig.~\ref{angprof}).  In particular, the profile in the Eastern region
is significantly higher, flatter and more complex than in the
complementary Western sector, being most extended in the
40\degr-80\degr\ angular region.  This is more pronounced at energies below
1.5 keV,  and most in the 0.9-1.5 keV band, which shows the greater extension
and flatter overall distribution.

\subsection{Individual point sources}
\label{individual}

\begin{figure}
\psfig{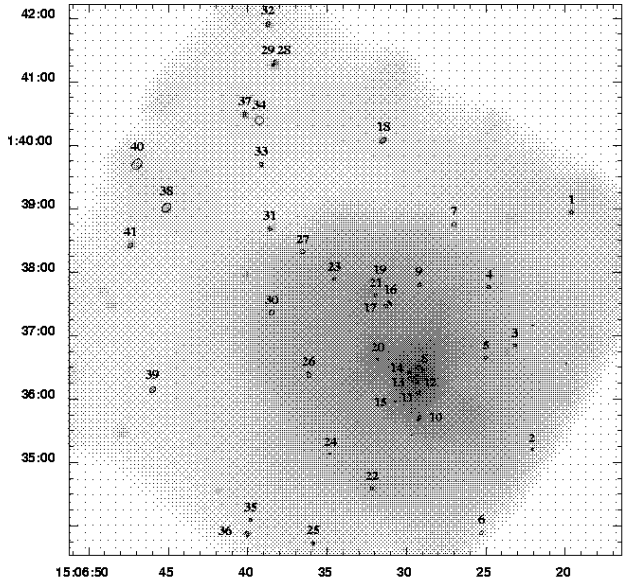}
\caption{The positions of the 41 sources are shown, together with the
size and shape of the region used to determine their count rate.  
Sources are in order of increasing right ascension.
}
\label{sources}
\end{figure}

The map in Fig.~\ref{large} indicates the presence of several individual
point-like 
sources scattered over the full field of view, as listed in
Table~\ref{sourcetab}.  The source list was derived combining the results
of different source detection algorithms provided by the CIAO software
({\tt celldetect, wavedetect}) on the full energy band image, and
on the two separate broad-band images used in Fig.~\ref{large}.  The
combined list was cleaned of duplicates and compared with
the adaptively smoothed images.  A few sources that were evident from
the map but were not included by any of the detection algorithms used
were added.  Net 
source counts were computed first 
in the full 0.3-10 keV energy range, assuming as
background the annular region at the same distance from the X-ray peak
as the source, cleaned of the actual sources.  
The annuli chosen for this purpose are: $0\farcm5-0\farcm75$, $0\farcm75-1'$,
$1'-2'$, $2'-4'$ and $>4'$.  Within $0\farcm5$, both algorithms find many
overlapping sources, grouped in a sort of elongated pattern in two main
regions.  We have substituted the many ``individual" sources with two ellipses
comprising the combined regions of the many individual sources, which we will
consider later.  In any case, the region within $0\farcm5$ is not 
considered at this stage.  The list was further cleaned of
sources with a signal-to-noise smaller than $\sim$ 3.   Source regions and
their numbers (in increasing RA) are shown in Fig.~\ref{sources},
superposed onto the adaptively smoothed image.  As can be
seen from the plot, most peaks evident in the adaptively smoothed image are
accounted for, while a few others are not included in the final list, because
their signal-to-noise in the total band is smaller than our set threshold.

Since most of the detected sources have very
few counts, attempts at determining fluxes or count rates in different
energy bands have resulted in only a few detections.  We therefore do
not calculate hardness ratios for our detected sources.   However,
we notice that the distribution of the counts in broad energy bands
could be very different in different sources (see also the discussion on
hardness ratios in Blanton et al.\ 2000 and Sarazin et al.\ 2001). 

\begin{table*}
\caption[]{Positions, net counts in the broad energy band (0.3-10 keV),
fluxes and luminosities (0.3-10 keV) of the sources detected in NGC
5846.  Sources inside of a radius r=$0\farcm5$ from the center are large
regions (ellipses) along and above structures evident in the X-ray maps
(see \S~\ref{individual} for details), and are within parenthesis in
the table.  Net counts are
converted into fluxes assuming a conversion factor 2.6$\times 10^{-16}$
erg cm$^{-2}$ s$^{-1}$ counts$^{-1}$. 
}
\label{sourcetab}
\begin{flushleft}
\scriptsize{
\begin{tabular}{lllrrrrr}
\hline
\hline
Source&\multicolumn{1}{c}{R.A.}&\multicolumn{1}{c}{Dec.}&Total &Net &
Error&\multicolumn{1}{c}{f$_x$}&\multicolumn{1}{c}{L$_x$} \\
Number&\multicolumn{1}{c}{(J2000)}&\multicolumn{1}{c}{(J2000)}&\multicolumn{1}
{c}{Counts}&\multicolumn{1}{c}{Counts}&&\multicolumn{1}{c}{(0.3-10 keV)}
\\
\hline\hline
1 & 15:06:19.601 & +01:38:56.73& 15 & 14.1 & 3.8 & 3.7$\times
10^{15}$ & 4.5$\times 10^{38 }$\\
2 & 15:06:22.098 & +01:35:12.50& 10 & 9.8 & 3.2 & 2.6$\times
10^{15}$ & 3.1$\times 10^{38 }$\\
3 & 15:06:23.170 & +01:36:50.73& 14 & 12.1 & 3.7 & 3.1$\times
10^{15}$ & 3.9$\times 10^{38 }$\\
4 & 15:06:24.808 & +01:37:45.98& 54 & 51.4 & 7.3 & 1.34$\times
10^{14}$ & 1.64$\times 10^{39 }$\\
5 & 15:06:24.991 & +01:36:39.11& 15 & 12.4 & 3.8 & 3.2$\times
10^{15}$ & 3.9$\times 10^{38 }$\\
6 & 15:06:25.284 & +01:33:53.43& 15 & 13.5 & 3.8 & 3.5$\times
10^{15}$ & 4.3$\times 10^{38 }$\\
7 & 15:06:27.008 & +01:38:45.40& 24 & 22.1 & 4.9 & 5.8$\times
10^{15}$ & 7.1$\times 10^{38 }$\\
\llap{(}8 & 15:06:29.148 & +01:36:28.66& 393 & 291.9 & 19.9 &
7.59$\times 10^{14}$ & 9.32$\times 10^{39} \rlap{)}$\\
9 & 15:06:29.191 & +01:37:47.92& 17 & 13.0 & 4.1 & 3.4$\times
10^{15}$ & 4.2$\times 10^{38 }$\\
10 & 15:06:29.202 & +01:35:41.87& 47 & 35.1 & 6.9 & 9.1$\times
10^{15}$ & 1.12$\times 10^{39 }$\\
\llap{(}11 & 15:06:29.247 & +01:36:06.46& 101 & 55.3 & 10.1 &
1.44$\times 10^{14}$ & 1.77$\times 10^{39 }$\rlap{)}\\
\llap{(}12 & 15:06:29.335 & +01:36:20.76& 173 & 136.9 & 13.2 &
3.56$\times 10^{14}$ & 4.37$\times 10^{39 }$\\
\llap{(}13 & 15:06:29.549 & +01:36:17.87& 304 & 221.4 & 17.5 &
5.76$\times 10^{14}$ & 7.07$\times 10^{39 }$\rlap{)}\\
14 & 15:06:29.825 & +01:36:25.24& 40 & 24.8 & 6.3 & 6.4$\times
10^{15}$ & 7.9$\times 10^{38 }$\\
15 & 15:06:30.676 & +01:35:57.78& 15 & 13.2 & 3.9 & 3.4$\times
10^{15}$ & 4.2$\times 10^{38 }$\\
16 & 15:06:31.046 & +01:37:30.45& 41 & 38.2 & 6.4 & 9.9$\times
10^{15}$ & 1.22$\times 10^{39 }$\\
17 & 15:06:31.307 & +01:37:28.10& 24 & 20.2 & 4.9 & 5.3$\times
10^{15}$ & 6.5$\times 10^{38 }$\\
18 & 15:06:31.477 & +01:40:04.38& 21 & 18.5 & 4.6 & 4.8$\times
10^{15} $& 5.9$\times 10^{38 }$\\
29 & 15:06:31.702 & +01:37:51.45& 15 & 14.6 & 3.9 & 3.8$\times
10^{15}$ & 4.7$\times 10^{38 }$\\
20 & 15:06:31.820 & +01:36:37.72& 30 & 27.0 & 5.5 & 7.0$\times
10^{15} $& 8.6$\times 10^{38 }$\\
21 & 15:06:31.948 & +01:37:37.90& 22 & 18.9 & 4.7 & 4.9$\times
10^{15} $& 6.1$\times 10^{38 }$\\
22 & 15:06:32.164 & +01:34:35.90& 20 & 18.5 & 4.5 & 4.8$\times
10^{15}$ & 5.9$\times 10^{38 }$\\
23 & 15:06:34.563 & +01:37:53.60& 18 & 15.7 & 4.2 & 4.1$\times
10^{15} $& 5.0$\times 10^{38 }$\\
24 & 15:06:34.811 & +01:35:08.60& 25 & 24.5 & 5.0 & 6.4$\times
10^{15}$ & 7.8$\times 10^{38 }$\\
25 & 15:06:35.892 & +01:33:43.75& 25 & 23.8 & 5.0 & 6.2$\times
10^{15}$ & 7.6$\times 10^{38 }$\\
26 & 15:06:36.147 & +01:36:22.99& 21 & 14.9 & 4.6 & 3.9$\times
10^{15}$ & 4.7$\times 10^{38 }$\\
27 & 15:06:36.519 & +01:38:19.26& 15 & 13.1 & 3.9 & 3.4$\times
10^{15}$ & 4.2$\times 10^{38 }$\\
28 & 15:06:38.241 & +01:41:18.53& 13 & 12.2 & 3.6 & 3.2$\times
10^{15}$ & 3.9$\times 10^{38 }$\\
29 & 15:06:38.384 & +01:41:16.19& 16 & 15.6 & 4.0 & 4.0$\times
10^{15}$ & 5.0$\times 10^{38 }$\\
30 & 15:06:38.478 & +01:37:21.57& 13 & 10.8 & 3.6 & 2.8$\times
10^{15}$ & 3.4$\times 10^{38 }$\\
31 & 15:06:38.588 & +01:38:40.97& 16 & 14.8 & 4.0 & 3.9$\times
10^{15}$ & 4.7$\times 10^{38 }$\\
32 & 15:06:38.721 & +01:41:54.74& 16 & 14.9 & 4.0 & 3.9$\times
10^{15}$ & 4.8$\times 10^{38 }$\\
33 & 15:06:39.153 & +01:39:42.04& 12 & 11.0 & 3.5 & 2.9$\times
10^{15}$ & 3.5$\times 10^{38 }$\\
34 & 15:06:39.186 & +01:40:23.62& 22 & 17.1 & 4.7 & 4.4$\times
10^{15}$ & 5.5$\times 10^{38 }$\\
35 & 15:06:39.823 & +01:34:05.58& 11 & 10.0 & 3.3 & 2.6$\times
10^{15}$ & 3.2$\times 10^{38 }$\\
36 & 15:06:40.031 & +01:33:52.74& 23 & 21.5 & 4.8 & 5.6$\times
10^{15}$ & 6.9$\times 10^{38 }$\\
37 & 15:06:40.194 & +01:40:29.55& 20 & 19.2 & 4.5 & 5.0 $\times
10^{15}$ & 6.1$\times 10^{38 }$\\
38 & 15:06:45.118 & +01:39:00.99& 18 & 12.5 & 4.2 & 3.3$\times
10^{15}$ & 4.0 $\times 10^{38 }$\\
39 & 15:06:45.989 & +01:36:08.96& 16 & 13.7 & 4.0 & 3.6$\times
10^{15}$ & 4.4$\times 10^{38 }$\\
40 & 15:06:46.971 & +01:39:41.95& 25 & 17.9 & 5.0 & 4.7$\times
10^{15}$ & 5.7$\times 10^{38 }$\\
41 & 15:06:47.386 & +01:38:25.02& 11 & 9.6 & 3.3 & 2.5$\times
10^{15}$ & 3.1$\times 10^{38 }$\\
\hline \hline 
\end{tabular}  
}
\end{flushleft} 
\end{table*}


\begin{figure}
\psfig{file=2278.f6,width=9cm,clip=}
\caption{The luminosity distribution of the sources detected in the total
(0.3-10 keV) energy band.
}
\label{ldistr}
\end{figure}

Count rates in the total energy band are converted into fluxes 
assuming a constant conversion rate of 1 count s$^{-1}$=
7$\times 10^{-12}$ erg cm$^{-2}$ s$^{-1}$ 
(see Table~\ref{sources}).  The X-ray luminosities (Fig~\ref{ldistr}) are
distributed in the range $\sim 3\times 10^{38}
 \la L_x \la 2\times 10^{39}$ \mbox{erg 
s$^{-1}$}.  Four sources in the inner $30''$ region (\# 8, 11, 12, 13) 
have not been included in the analysis of individual sources since they
represent extended regions as discussed above.  The total number of
``individual" sources therefore is 37.  Although we do not claim
completeness at the lower luminosities (see later), we should have
detected all of the bright sources, and we have detected 
about 30 sources in the central 4$'$ radius of NGC 5846, which most
likely all belong to the galaxy.

\begin{figure*}
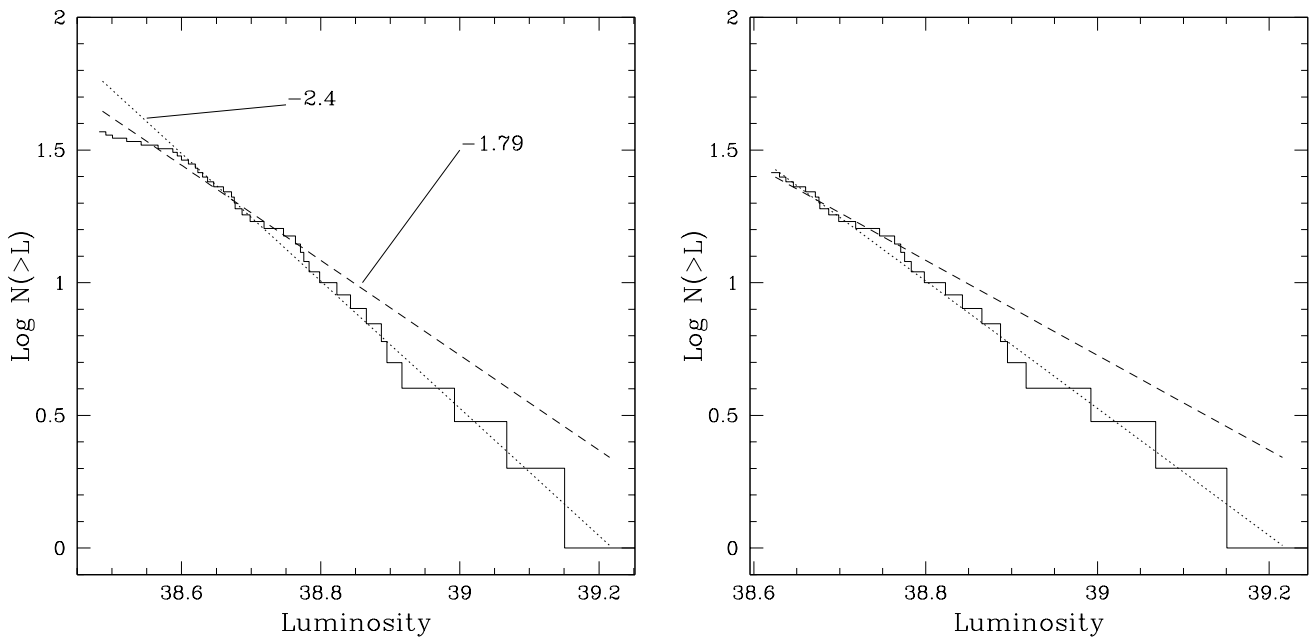

\resizebox{18cm}{!}{
\psfig{file=2278.f7a,width=9cm,clip=}
\psfig{file=2278.f7b,width=9cm,clip=}
}
\caption{The luminosity function of the sources in NGC 5846
(0.3-10 keV energy band).  The dashed slope represents the value found by
Shirey et al.\ (2001) for the M31 sources, although at lower L$_x$,
normalized to the present data.  
The -2.4 value appears to better represent the
NGC 5846 data than the flatter function. 
Left panel: all sources listed in Table~\ref{sources}. Right panel: only
sources with a Signal-to-noise ratio greater than 3.4.
}
\label{lumfun}
\end{figure*}

Figure~\ref{lumfun} shows the luminosity function of the sources
detected here.  We have assumed that all are associated with 
NGC~5846,  although we
expect that a few of them might be (mostly background) intruders ($< 9$
for a flux limit f$_{0.5-2}\sim 2\times 10^{-15}$ \ergsec\ in the whole
field of view, Hasinger et al. 1998).
A power law distribution is suggested for luminosities
above $\sim 4 \times 10^{38}$ \ergsec, with a turnover at lower
luminosities.  Although breaks in the luminosity functions of NGC 1553
and NGC 4697 have been suggested at similar luminosities (Blanton et
al.\ 2000; Sarazin et al.\  2000), in NGC 5846 sources at low
luminosities (low count rates) are detected with less than or about 10 counts,
and are already in most cases at the detection limit, so we cannot
exclude a large bias due to incompleteness.  
The right panel in fact shows the luminosity function derived for sources
detected at higher significance (S/N $>$ 3.4).  The flattening at low
luminosity has disappeared almost completely. 
Also, 
the flattening becomes 
{\it more\/} conspicuous when sources of lower significance (2.5 $\sigma$) are
included (plot not shown). 

The dashed line drawn over the steeper part of the luminosity 
function represents the slope ($-1.79$) 
derived for the high-luminosity end of the luminosity function in M31
(Shirey et al.\ 2001), which is the steepest value reported so far in the
literature for early-type galaxies or bulges. In NGC 5846,  an even steeper
value of $-$2.4 might be more appropriate to cover the full range of
luminosities sampled.   
Since the luminosity range sampled for NGC 
1553, NGC 4697, and NGC 5846 is very similar, this difference in slopes must 
indicate some other phenomenon at play.

\begin{figure}
\psfig{file=2278.f8,width=9cm,clip=}
\caption{The energy distribution of the photons in 4 regions within
a radius of $0\farcm75$ at the galaxy center.  All distributions are
normalized to 1 at their peak, and are binned in the same energy bins 
for easy comparison.  Filled circles and triangles 
are for the innermost
regions, representative of the higher and lower surface brightness
regions at the center, 
respectively. 
Open squares and open circles refer to regions at 
distances $\sim 0\farcm5$ S and ENE of the center, respectively.  Horizontal
error bars indicate the size of the energy bins. Points are plotted
offset from the center of the bin to avoid confusion. 
The dotted line indicates the average energy distribution of the 
photons in the $0\farcm5 - 0\farcm75$ region.  A finer 
energy 
binning is used for this region. Point sources detected at 2.5 $\sigma$ and
above have been masked out from all regions.
}
\label{endist}
\end{figure}

\subsection{Energy distribution}
\label{energydist}
Given the complexity of the spectral characteristics of the extended
emission, as already 
shown by the maps in Fig.~\ref{manye}, we defer a 
more complete
treatment of the spectral data to a later paper.  
However, it should be 
noticed that the spectral characteristics of the gas are at least as
complex as its morphological distribution.

Fig.~\ref{endist} better illustrates the complexity of the emission on
small scales.  The energy distributions of the photons in four of the
regions analyzed are compared. All regions selected are within a
$0\farcm75$  radius from the center of the galaxy, and cover an area
between $\sim$ 50 arcsec$^{2}$ and $\sim$ 300 arcsec$^{2}$.  The data
are binned in relatively large bins, to obtain a good statistical
significance at each point and within the same energy boundaries,
albeit at the expense of 
details.\footnote{The spectral matrixes for each
region do not take into account the extended nature of the sources.
While formally incorrect, the errors introduced by this assumption are
very small on the scale of the sources analyzed here, and are also not
very significant throughout the S3 chip (J. McDowell,
private communication, see also Hicks et al. 2002).}
the very center of the galaxy, covering the higher surface brightness
structure and its adjacent lower surface brightness region.  The other two
are at slightly larger radial distance and cover the ENE and the S
enhancements visible in the maps (Fig.~\ref{manye}).  It is evident
from the figure that the distributions are not the same, both in shape
and in the position of the peak.  The average distribution at the same
radial distance also appears to have a broader shape than for example
the S region (open squares).

\begin{figure}
\psfig{file=2278.f9a,width=9cm,angle=-90,clip=}
\psfig{file=2278.f9b,width=9cm,angle=-90,clip=}
\caption{The energy distribution of the photons in 3 adjacent annuli
with ``inner$-$outer'' radii of $45''-60''$, $1'-2'$ and $2'-4'$, and in 2
regions of higher and lower surface brightness within the inner 30$''$.  
Detected point sources have been masked out. 
The background is obtained from the easternmost corner of the CCD field
(see text) and subtracted in the outer annuli (top panel)
normalized by the respective areas.  
The net counts are
binned to retain a 2.5--3$\sigma$ significance in each bin.
}
\label{specdis}
\end{figure}

Even more dramatic is the comparison of the photon distribution on
larger areas and at large
radii. 
The two lower panels of 
Figure~\ref{specdis} show the comparison between
two inner regions, characterized by higher and lower surface
brightness (the former being along the high surface brightness feature,
inclusive of sources 8, 12, 13;  the latter in the lower surface
brightness region to the ENE of it).
The upper panel of Figure~\ref{specdis} shows the energy distribution in
three outer concentric annuli. 
These are defined as having inner/outer radii
of $0\farcm75-1';~1'-2';~2'-4'$ respectively; all detected
sources have been removed.  
The background is estimated from the easternmost corner of the CCD, at
galactocentric radii of $>4'$ and subtracted from the data;
for all annuli care has been
taken to ensure that the area not covered by the CCD is also masked out.
The net counts have then been binned to 
obtain 
a $\ge 2.5 \sigma$ significance in each bin.  

\begin{table*}
\caption[]{Results of spectral fitting using the XSPEC package.
Counts are obtained in different regions
as described.  A combination of a plasma code (MEKAL) and Power law
(with $\Gamma$ fixed at a value of 1.84, best fit from the $1'-2'$
region) models is used. Errors on kT (fifth column) represent the 90\% 
probability intervals. DOF (sixth column) stands for Degrees of
Freedom.}
\label{specresults}
\begin{flushleft}\bigskip
\begin{tabular}{llrcll}
\hline
\hline
Region&Counts& Energy& N$_H$& kT& $\chi^2_\nu$ \\
   & $\pm$ error     &(keV) &  ($10^{-20}$) & error &(DOF) \\
\hline\hline
High surface &1652  &0.3-1.9 &6.6& 0.62&0.8 \\
brightness &   $\pm$41     & &&0.59-0.65&(36)\\
within 30$''$ \\
\dotfill \\
Low surface & 1865 &0.3-2.3& 6.3& 0.59 &1.2  \\
brightness  &  $\pm$43&& &0.57-0.62 & (62) \\
within 30$''$ \\
\dotfill \\
Ann 0\farcm75-1$'$ &3474& 0.3-2.4 &12& 0.60 &1.00  \\
   & $\pm$62 && &0.58-0.61&(54)\\
\dotfill \\
Ann $1'-2'$ &10872& 0.4-2.6& 14& 0.67  & 1.4  \\
 &$\pm$11&&&0.65-0.68& (98) \\
\dotfill \\
Ann $2'-4'$ & 5788 & 0.4-1.8&  15& 0.81  & 1.0  \\
 &$\pm$102&&&0.79-0.84& (51) \\
\hline \hline
\end{tabular}

\medskip
\end{flushleft}

Note:  There is no requirement of a Power law component in the $2'-4'$
region.
\end{table*}

It is apparent from the comparison that all regions have 
significantly different overall spectral shapes, and different positions of
the peak.  We have tried to fit the data with current spectral models,
as summarized in Table~\ref{specresults}.  
We stress that the results presented here are neither final nor the
best possible, but their purpose is to give a 
simplified parameterization of the average gas properties.
We have used 
a combination of plasma line code and simple thermal or 
power law models.  We find that {\it (i)\/} reasonable fits require
2 components out to 2$'$ radius.  Either a power law or Bremsstrahlung
are needed to account for a residual high energy tail and extreme low energy
absorption that a fit with only a plasma code required for the low
energy range would produce.
{\it (ii)\/} There is no strong evidence
that one model plasma code should be preferred. We used models based on 
both the Raymond-Smith code (hereafter {\it Raymond\/} code; Raymond \& Smith 
1977) and the MEKAL code (e.g., Kaastra et al.\ 1996). The MEKAL code includes 
a better treatment of the Iron L lines near 1 keV according to recent 
measurements and calculations of atomic parameters (Liedahl et al.\ 1995). The 
MEKAL code indeed yielded slightly better fits than the {\it Raymond\/} code 
(see, e.g., Fig.~\ref{specex}), but the required N$_H$ 
is significantly higher than for the {\it
Raymond\/} model, and than the line-of-sight value.  
{\it (iii)\/} The
fits are not sensitive to the abundance assumed; {\it (iv)\/} we find
significant residuals at $\sim 1.8$ keV and below $\sim 1$ keV (see
Fig.~\ref{specex}), no matter what combination of models and parameters
we use.  We have used the latest calibration provided by the ASC/CXC,
which significantly improve the quality of the fits over previous
versions; however, we cannot exclude residual instrument calibration
uncertainties.

Leaving both kT and $\Gamma$ as free parameters, 
the best fit temperatures derived from the
MEKAL model are not significantly different among the regions considered
within 2$'$, in spite
of the different photon distribution shown in Fig.~\ref{specdis}.  
The power law slopes instead have very different best fit values
($\Gamma$ in the range 0.15-2.5), 
however with large errors.  
If we fix the power law slope to a common value (of $\Gamma$=1.85, best
fit value from the $1'-2'$ region), we find that the temperature of the
plasma increases towards the outside, from the low surface brightness
central feature to the $2'-4'$ annulus.  The low and high surface
brightness features have formally different best fit temperatures,
however the difference is not statistically significant (see
Table~\ref{specresults}).   The N$_H$ parameter is consistent with the
line of sight value only in the very inner regions.

\begin{figure}
\psfig{file=2278.f10a,width=9cm,angle=-90,clip=}
\psfig{file=2278.f10b,width=9cm,angle=-90,clip=}
\caption{Spectrum of the $1'-2'$ annulus, fitted with 1 T plasma code
and a power law.  The best fit parameters are:  top panel:
N$_H$=1.4$\times 10^{21}$ cm$^{-2}$; kT(MEKAL)=0.66; $\Gamma$(PL)=1.8;
$\chi^{2}_\nu$=1.4; bottom panel:
N$_H$=7.5$\times 10^{20}$ cm$^{-2}$; kT({\it Raymond\/})=0.78;
$\Gamma$(PL)=1.7; $\chi^{2}_\nu$=1.7. 
Abundances are in both cases $ 1\times $ solar. 
Data are fitted in the 0.4-2.6 keV range. Local background is
subtracted. 
}
\label{specex}
\end{figure}

\section{Discussion}
\label{discussion}
The complexity of the emission in NGC 5846 is self-evident from the maps
and from the comparison of the images at different energies.  The high-energy
band, above $\sim 3$ keV, is completely dominated by a collection 
of individual sources that are now becoming a common feature in early
type galaxies. In the soft energies, a very complex morphology is
observed within the central $1'-2'$, with accompanying complexity in the
gas physical and chemical properties, while a more regular 
(i.e., diffuse) 
gas distribution is observed at larger radii.  

\subsection{Individual sources}

As already suggested by the crude properties of the low-luminosity
early-type galaxies observed with {\it Einstein\/} (Fabbiano, Kim \&
Trinchieri 1992), and then measured with ASCA and BeppoSAX (Matsushida
et al.\ 1994; Matsumoto et al.\ 1997; Pellegrini 1999; 
Trinchieri et al.\ 2000), a population of individual
sources, most likely low mass X-ray binaries, constitutes the dominant
source of emission at high energies in ``normal" early-type galaxies of
all luminosities.  A thorough study of this population is now possible
with {\it Chandra}, due to its high spatial resolution and sensitivity.  An
example of such detailed study in the low-luminosity elliptical NGC 4697
is presented by Sarazin et al.\ (2000, 2001) where $\sim 90$ sources with
luminosities above $\sim 5 \times 10^{37}$ erg s$^{-1}$ are detected.
A smaller though substantial population is also measured in NGC
1553 (Blanton et al.\ 2000), with $\sim 40$ sources above $\sim 3 \times
10^{38}$ \ergsec.

We detected $\sim$\,40 sources 
in NGC 5846, all with luminosities above $\sim 3\times 10^{38}$ \ergsec,
which is higher than the Eddington luminosity for a 1 M$_\odot$ accreting
object.  Ultra-Luminous Sources (ULS) have been known to exist in spiral
galaxies for quite some time (Fabbiano 1989) but more examples of these 
non nuclear bright sources are becoming available 
with \chandra\ data in both spiral (e.g. 
Tennant et al.\ 2001, Prestwich 2001) and elliptical galaxies (e.g. Blanton
et al.\ 2000 and Sarazin et al.\ 2001). 
The brightest sources (L$_X >2\times 10^{39}$ \ergsec) so far have
almost esclusively been associated with spiral galaxies and 
regions of star formation
activity  (e.g., Zezas, Georgantopoulos \& Ward 1999; Roberts \& Warwick
2000; Fabbiano, Zezas \& Murray 2001), and are known to vary both in
intensity and in spectral characteristics (Mizuno, Kubota \& Makishima
2001), supporting the idea that at least most of them 
are powered by a black hole.   If the accretion is isotropic, these ULS's 
would require intermediate-mass black holes, which are however
inconsistent with measured inner disk temperatures (Kubota et al. 2001),
and either anisotropic or beamed emission have been proposed to explain
their high luminosities (King et al. 2001; K\"ording, Falcke \& Markoff 2002).

Although not as extreme, several
point sources  emitting above the Eddington luminosity are now also
found in ``normal" early-type galaxies, in locations without any
association with star formation activity. 
A recent discussion on binary sources in early-type galaxies
associates most of them, 
in particular 
the brighter ones, with the globular cluster population.  In NGC 1399,
Angelini et al.\ (2001) remark that the 
average luminosity of the globular cluster sources is higher than the
rest of them. A high percentage of the sources in both NGC 4697 and NGC
1553 are also associated with globular clusters.  
As these three galaxies cover a large range in $L_X/L_B$ ratio, it seems
likely 
that a large fraction of X-ray sources is associated with the globular
cluster population in {\it all\/} early-type galaxies. 
It is quite possible that the same association exists for sources in
NGC 5846, which has a rich population of them (Forbes et al.\ 1997
detect over 1200 globular clusters using WFPC2 images from HST).
However, this is not at all what is observed in our Galaxy or M\,31,
where {\it no\/} sources brighter than the Eddington limit are found
associated with Globular clusters.

The comparison of the luminosity distributions of the sources in the
three early-type galaxies available (data on NGC 1553 and NGC 4697 from the
literature) indicates a lack of high-luminosity sources in NGC 5846
relative to the other two (i.e., a steeper slope).  We cannot confirm a
flattening in the luminosity function at low luminosities, which is
most likely an effect of incompleteness.  
NGC 5846 is $\sim$\,50\% more distant than NGC 1553 and $\sim$\,100\% more
than NGC 4697, the observing time on NGC~5846 was somewhat shorter than for
the other two galaxies, and a much more significant component of
hot gas is present in NGC 5846, therefore our limiting threshold is
higher ($\sim 6\times$ than for NGC 4697, and only marginally than for 
NGC~1553), so we are not sampling the lower luminosity range as well as
in these two galaxies. 
Furthermore, given the larger distance, our source confusion may be more
severe.
We expect this to be most relevant at the lower end of the luminosity
distribution (brighter sources being rarer, and their luminosities being
dominant over the addition of smaller contributions), which could
artificially populate the low-to-intermediate luminosity range considered,
resulting in a steeper relation.  We therefore cannot claim
that a break exists 
in the luminosity function of NGC~5846, since the flattening observed below
$L_X \sim 4 \times 10^{38}$ \ergsec is almost surely due to
incompleteness at the low end of the $L_X$ distribution.
However, we detect a steeper slope above $\sim 6 \times
10^{38}$ \ergsec\ relative to the luminosity functions of NGC~1553 and
NGC~4697.  

\subsection{Large scale distribution of the hot gas}

Given the small field of view of the S3
detector relative to the full spatial extent of the X-ray emission in this
object, little can be said about the large-scale structure of the gas.
In particular, the W quadrant is only covered out to $r \sim 2'-3'$,
before reaching the edges of the CCD.  To the E, the coverage is
better, but only out to $r \sim 4'-5'$ for a large fraction of the
galaxy.

As shown by the radial profiles in Fig.~\ref{rawprof} and Fig.~\ref{angprof},
extended emission is detected out to $\ga 6'$, i.e., to the boundary of
the CCD field of view.  The shape of the photon distribution appears
broader above than below $\sim 1$ keV, and shows azimuthal asymmetries
at all energies.  In particular, a strong enhancement is observed in
the 0\degr $<$ PA $<$ 120\degr\ region, between $\sim 1'-2\farcm5$ radius,
more pronounced below 1 keV.  Moreover, the 0\degr $<$ PA $<$ 90\degr\ sector
profile appears flatter than the remaining sectors, mostly below $\sim 1$ keV.
The 
significant 
azimuthal asymmetries in the photon distribution prevents us
from parameterizing the 
radial profile 
of the emission with a ``King-type'' profile as done in the past.  While this
kind of profile might still be valid at larger radii, the inner region
clearly requires a much more complex model. 

The spectral data also indicate that a complex spectral model is
needed, which appears to vary  even within relatively small regions.
At large radii (r$>2'$) only the eastern portion of the galaxy is
covered, and the spectral photon distribution can be fitted reasonably
well with a single-temperature plasma with kT$\sim$ of 0.8 keV.  It is
likely that at these radii the turmoil affecting the innermost parts of
the galaxy is less severe and the gas is in 
an energetically 
stabler configuration. However, the lack of coverage at larger radii prevents
us from a more complete analysis of these regions.

\begin{figure*}
\resizebox{18cm}{!}
{
\psfig{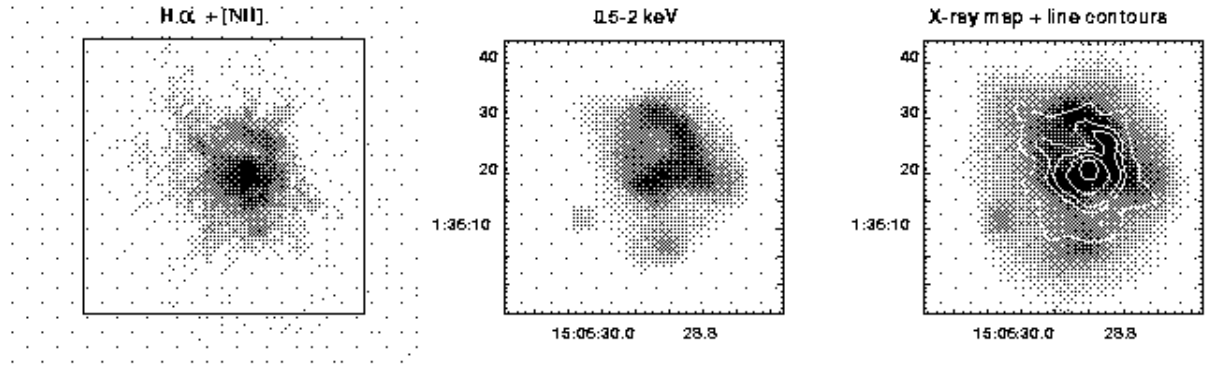}
}
\caption{
Comparison of the innermost region of NGC 5846 in the H$\alpha$+[N\,{\sc ii}]
lines and in X rays (0.5-2.0 keV). 
The data are displayed on the same scale. H$\alpha$+[N\,{\sc ii}] contours on
the X-ray smoothed image are shown in the right panel.
H$\alpha$+[N\,{\sc ii}] line filter images were obtained with EFOSC1 at the
ESO 3.6-m telescope (see Trinchieri \& di Serego Alighieri 1991).  
}
\label{xha}
\end{figure*}

\subsection{Small scale properties of the hot gas}

\begin{figure*}
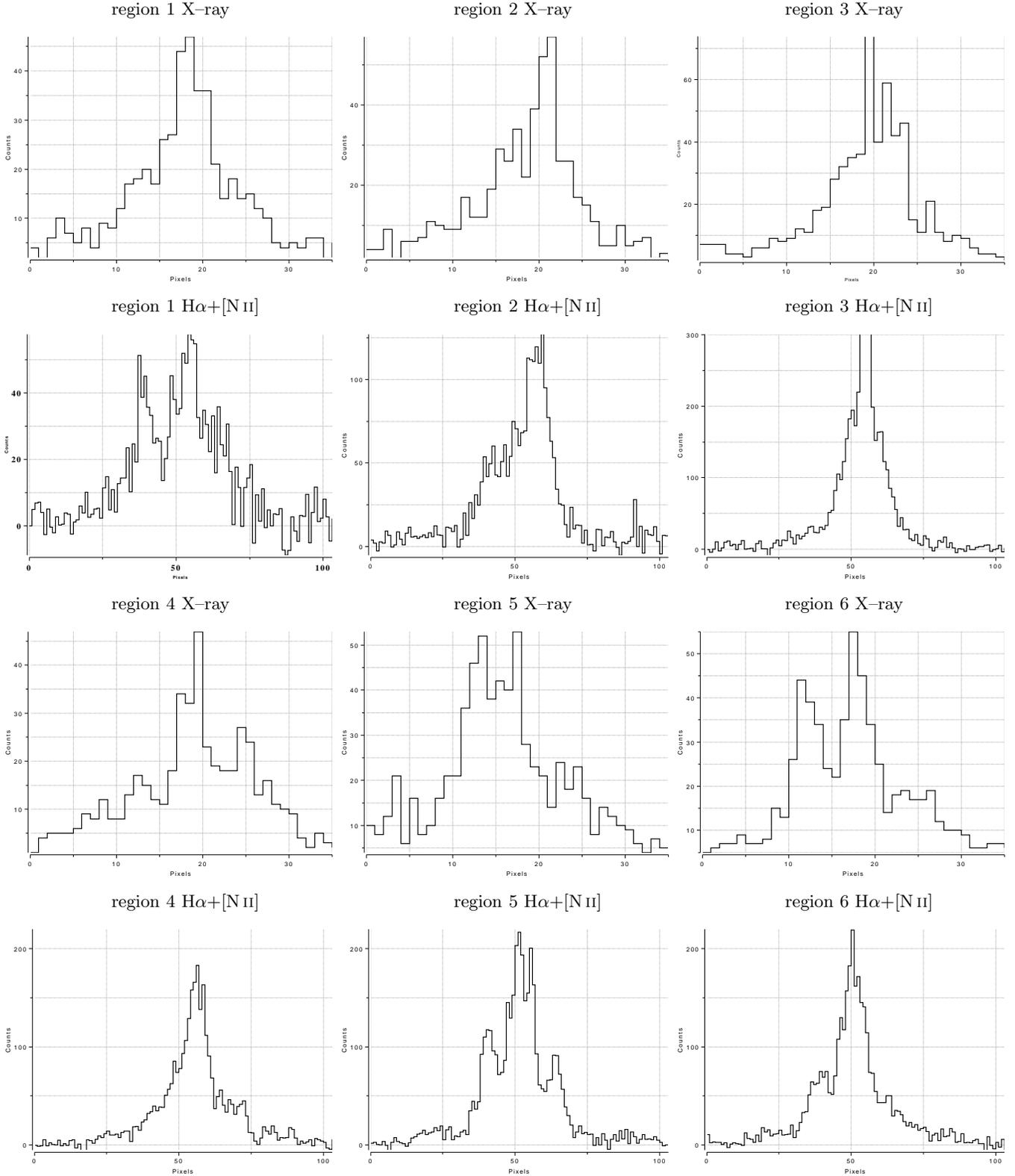

\unitlength1.0cm
\begin{picture}(18,21)
\thicklines
\put(0,0.0){
\begin{picture}(18,9.0)
\resizebox{18cm}{!}
{
\psfig{figure=2278.f12a,width=18cm,clip=}
\psfig{figure=2278.f12b,width=18cm,clip=}
\psfig{figure=2278.f12c,width=18cm,clip=}
}
\end{picture}}
\put(2.0,4.8){
region 4  H$\alpha$+[N\,{\sc ii}]}
\put(8.1,4.8) {
region 5  H$\alpha$+[N\,{\sc ii}]}
\put(14,4.8){
region 6 H$\alpha$+[N\,{\sc ii}]
}
\put(0,5.3){
\begin{picture}(18,9.0)
\resizebox{18cm}{!}
{
\psfig{figure=2278.f12d,width=18cm,clip=}
\psfig{figure=2278.f12e,width=18cm,clip=}
\psfig{figure=2278.f12f,width=18cm,clip=}
}
\end{picture}}
\put(2.0,10.1){
region 4 X--ray}
\put(8.1,10.1){
region 5 X--ray}
\put(14,10.1){
region 6 X--ray
}
\put(0,10.6){
\begin{picture}(18,9.0)
\resizebox{18cm}{!}
{
\psfig{figure=2278.f12g,width=18cm,clip=}
\psfig{figure=2278.f12h,width=18cm,clip=}
\psfig{figure=2278.f12i,width=18cm,clip=}
}
\end{picture}}
\put(2.0,15.4){
region 1 H$\alpha$+[N\,{\sc ii}]}
\put(8.1,15.4){
region 2 H$\alpha$+[N\,{\sc ii}]}
\put(14,15.4){
region 3 H$\alpha$+[N\,{\sc ii}]
}
\put(0,15.9){
\begin{picture}(18,9.0)
\resizebox{18cm}{!}
{
\psfig{figure=2278.f12j,width=18cm,clip=}
\psfig{figure=2278.f12k,width=18cm,clip=}
\psfig{figure=2278.f12l,width=18cm,clip=}
}
\end{picture}}
\put(2.0,20.7){
region 1 X--ray}
\put(8.1,20.7){
region 2 X--ray}
\put(14,20.7){
region 3 X--ray
}
\end{picture}
\caption{
Cuts along the horizontal (1 through 4) and vertical (5 and 6) 
axes through the central regions of NGC~5846 in the X-ray (first and third
row) and the H$\alpha$+[N\,{\sc ii}] (second and fourth row) images.
Horizontal scale is 70$''$, and the slits are positioned in the same place in
the two images.  Pixel sizes are 2$''$ in the
X-ray and $0\farcs675$ in the H$\alpha$+[N\,{\sc ii}] image.  Cuts are across
the high and low surface brightness central features as shown in
Fig.~\ref{cuts}. The sharp peak in the H$\alpha$ cut \# 3 has been set
off-scale to better show the lower part of the curve.  Vertical scales  could
not be set equal in all panels, 
due to the different dynamical ranges in different regions. 
}
\label{xhaproj}
\end{figure*}

The central region of NGC 5846 shows complex morphology in the form of
a prominent ``hook'' at the center, plus several ``blobs'' and filaments
to the NNE at $\sim 1'-3'$ radius and towards S,
that had already been seen in the ROSAT HRI images
(Trinchieri et al.\ 1997a).    The spectral properties
are not the same in different regions, suggesting 
a rather complex physical state 
of the interstellar medium.

Complex gas morphologies are now more common as \chandra\ data on
bright early-type galaxies become available.  NGC 1553 (Blanton et al.\ 
2000) and NGC 4636 (Jones et al.\ 2001) 
show a spiral-like feature at the center, NGC 4696 (Sanders \& Fabian 2001) 
exhibits
a spiral plume-like feature, and a disturbed inner halo is seen in NGC
1399 (Angelini et al.\ 2001), to name a few. 
In several objects, the peculiar morphologies are associated with 
central activity, in the form of an extended
radio source (e.g., NGC 1399) 
or of a previously unknown (obscured, or low-luminosity) X-ray central
source ($e.g.$ NGC 1553), but not in all (e.g., NGC 4636 has neither
an X-ray nucleus nor a strong radio source).  However, what makes NGC~5846 
galaxy unique (up to now!) relative to the other examples is the close
agreement found between the morphology in the X-ray gas and in the optical
line-emitting gas (H$\alpha$+[N\,{\sc ii}]) and dust.    
Several objects show complex line-emitting morphologies, but so far claims
of similarities do not extend to the very inner regions, like in NGC 5846
(see Fig.\ref{xha}).   
Even the morphological similarity found for NGC~1553 with the ROSAT HRI data
(Trinchieri et al.\ 1997a) does not appear to extend to smaller scales
(Blanton et al.\ 2000).   

Morphological agreement on arcmin scales was already suggested 
by the ROSAT HRI data (Trinchieri et al.\ 1997a; Goudfrooij \& Trinchieri
1998).  With the present data, the morphological similarity with the
optical line emission is brought an order of magnitude closer in,
to scales as small as a few arcsec (cf.\ Fig.~\ref{xha}), equivalent to a
few hundred parsecs, while no equivalent distortions are seen in the red
continuum image (Fig.~\ref{manye}).  Dust absorption, also detected in
optical images of NGC~5846 (Goudfrooij \& Trinchieri 1998; Forbes et al.\ 
1997), also displays a very similar filamentary structure extending down
towards the nucleus. 

The close similarity in the distributions of the X-ray and 
H$\alpha$+[N\,{\sc ii}] emission is rather stunning. In some aspects it is 
almost an exact replica in both morphology and intensity variations, in others 
at least the morphology is similar. For example, the ``two-horned'' H$\alpha$ 
structure is reflected in a stronger and longer ``horn'' in X-rays in the 
innermost crosscut (right ``horn''), while the left feature is visible as a 
distortion in the gas at larger radii.  Horizontal and vertical cuts at 
different locations through the X-ray and H$\alpha$+[N\,{\sc ii}] images 
confirm the visual impression of similar pattern at different intensities  
(Fig.~\ref{xhaproj}) and give a more quantitative view of the large intensity  
contrasts present at both wavelengths.   In crosscut 2, X-ray and 
H$\alpha$+[N\,{\sc ii}] emission are similar both in extent ($\le 30''$) and 
in relative intensities ($\sim \times 2$ from peak to shoulder), and a 
similarly broad base is present in crosscuts 3 and 4, although the very 
intense H$\alpha$ peak has no correspondence in the X-rays.  The comparison of 
crosscuts 1 and 6 would suggest that the panels have been exchanged, since the 
X-ray cut of region 1 closely resembles the H$\alpha$ cut of region 6 and 
vice-versa.  Finally, crosscut 5 appears somewhat more extended in X-rays, and 
with a larger plateau. 

What is the nature of 
these intermediate-- and small-scale morphological
structures?  The galaxy is at the center of a small group, and the
large-scale gas distribution shown by the PSPC data is regular and almost
azimuthally symmetric (Finoguenov et al.\ 1999). The galaxy is not
associated with a strong, extended radio source:\ A partially-resolved radio
source is detected at 1.4 and 8.4 GHz with the VLA (Filho, Barthel \& Ho
2000; M\"ollenhoff, Hummel \& Bender 1992), at a flux density that is several
orders of magnitude fainter than cores of more classical radio galaxies
(Fanaroff \& Riley 1974) and closer in power to the weak cores of Seyfert
galaxies.  We therefore do not expect it to cause a significant displacement
of the hot gas as observed in stronger, extended radio sources. We also do not
find any evidence for the presence of an X-ray-bright nuclear source which might
cause turmoil in the center. 

A ``high-speed encounter" has been postulated between NGC 5846 and its close
companion NGC 5850 (Higdon et al.\ 1998), mainly to explain the latter
galaxy's peculiar morphology.  This is not likely to explain the 
disturbed appearance in the hot gas distribution of NGC 5846, since the
effects should be visible also at larger distances and in particular
towards the spiral galaxy itself.  If anything, the X-ray contours
towards NGC 5850 (i.e., to the E) appear steeper on scales $> 5'$, which
might argue for gas compression if the galaxy is approaching (cf.\ Fig.\ 1 in
Finoguenov et al.\ 1999; note that Higdon et al.\ 1998 discard a ram-pressure 
stripping explanation for the morphology of NGC 5850, due to the lack of
extended X-ray emission in the system).  
At smaller scales, the distortion is towards the NNE, and its shape 
is more reminiscent of 
a tidal effect of an escaping object or a captured object spiraling in
towards the galaxy centre.  The existence of a 
similar spur in the morphology of the optical emission-line gas 
also suggest influence from an ``external source''.  
The kinematics of the H$\alpha$+[N\,{\sc ii}] emission is found to 
be  irregular, with steep velocity gradients in the the N--S
direction, and a sudden change of sign of velocity towards the S. 
Conversely, the kinematics of the stellar component is regular, showing slow
rotation (Caon et al.\ 2000). The kinematically distinct inner gas component 
supports the idea of external accretion, when the gas has not yet settled
into an equilibrium configuration.  
A similar interpretation was put forward by Sparks,
Macchetto \& Golombek (1989) and de Jong et al.\ (1990) for the nature of the
optical nebulosity in NGC 4696, which they associate with a
small, gas-rich 
galaxy that has fallen in and has been stripped of its gas and dust.  

Given the likely external origin of the optical nebulosity and dust in
NGC~5846, the idea of a causal connection between the dust and optical
nebulosity on one side and the hot gas on the other may seem
counterintuitive, since the hot gas most likely has an ``internal'' origin,
coming from mass loss of old stars (e.g., Fabbiano 1989). Sparks, Macchetto
\& Golombek (1989) and de Jong et al.\ (1990) independently suggested that
the enhancement of the X-ray emission associated with the colder gas and dust
is caused by electron conduction. The hot electrons 
thus act as cooling agent for the hot gas, while causing transient heating 
for dust grains and plausibly even acting as ionizing source for the optical
emission-line gas (see also Voit \& Donahue 1997; Goudfrooij \& Trinchieri
1998).  
This model was suggested to explain the larger-scale similarities in
NGC~5846 observed with the ROSAT HRI (Trinchieri et al.\ 1997a; Goudfrooij \&
Trinchieri 1998).  We can now give a quantitative estimate of the
expected luminosities in the features observed at small scales.

We have estimated an average X-ray surface brightness in the ``hook" of
$\sim 7$ counts arcmin$^{-2}$, corresponding to $\sim$ 1.4 $\times
10^{38}$ erg s$^{-1}$ pc$^{-2}$ at the adopted distance.  Although we do
not have a very precise assessment of the temperature of the gas, the
derived density depends only weakly 
on temperature. Hence we 
choose a uniform temperature of $kT$ = 0.6 keV. 
The choice of the assumed volume is however 
more critical, and we do not have direct information about the depth of
the structure.  Assuming that the emission comes from a narrow, bent
ellipsoid, with depth equivalent to the narrow side, the average
density of the hot gas is $n_e = 0.35$ cm$^{-3}$.  Following Macchetto
et al.\ (1996, as derived from Cowie \& McKee 1977), the predicted line
flux is L$\rm _{H\alpha}$ $\sim$2.4 $\times 10^{40}$
erg s$^{-1}$, assuming 1\% of the available energy is radiated in the
H$\alpha$+[N\,{\sc ii}] lines. This is to be compared to L$\rm _{H\alpha}$
$\sim$1.3 $\times 10^{40}$ erg s$^{-1}$ observed.  The good agreement between
the two estimates reinforces the likelihood of the ``heat conduction'' model.

\begin{figure}
\psfig{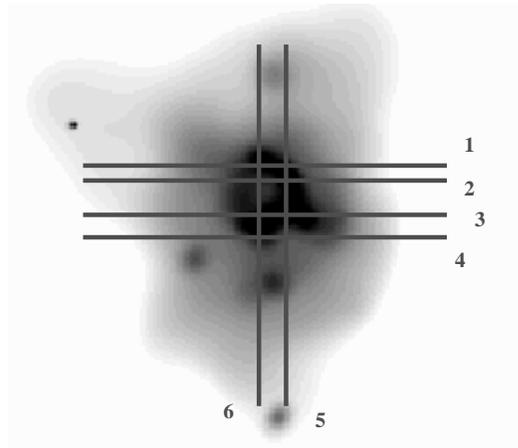}
\caption{
Position of the cuts used to derive the plots in Fig.~\ref{xhaproj}
}
\label{cuts}
\end{figure}

\section{Conclusions}

High resolution images of NGC 5846 obtained with \chandra\ show a disturbed
hot gas morphology on arcsecond scale ($1'' = 140$ pc) with unprecedented
details.   The morphological peculiarities are coupled with a complex
energy distribution of the photons.  Using these high-resolution
observations, the previously reported similarity of the X-ray emission and the
H$\alpha$+[N\,{\sc ii}] features are now extended into the very inner regions
of the galaxy, 
reinforcing the idea that there must be a close link between these two phases
of the ISM. A viable interpretation of this phenomenon is that the gas
ionization responsible for the optical line emission is provided by the 
electrons that also take care of the heat conduction within the hot gas.

About 40 point-like sources are found in NGC~5846, with $L_X > 3 \times
10^{38}$ \ergsec.  The luminosity function of these sources appears
steeper than any previously reported for early-type galaxies.  However, we
cannot exclude a bias at the low-luminosity end that might artificially
increase the contribution of low-luminosity sources.  We also cannot confirm
the change in slope observed at low luminosities in other early-type galaxies. 

\acknowledgements{We thank the whole {\it Chandra} team for help and
support during the data analysis and for many useful discussions on the
different aspects of the new data.  In particular we thank E. Mandel and  B.
Joye for their continuous support and good will to improve the
$funtools$/DS9 software, that we have heavily used in our analysis. 
This work has received partial financial support from the Italian Space
Agency. PG was affiliated with the Astrophysics Division of the 
Space Science Department of the European Space Agency during part of this 
project.}

\end{document}